\documentclass[
     final            % use final for the camera ready runs
]
  {aipproc}

\layoutstyle{6x9}
\usepackage{psfrag}

\begin{document}

\title{Center Vortex Model for the Infrared Sector of $SU(3)$ Yang-Mills
       Theory}


\author{Markus Quandt}{
  address={Institute for Theoretical Physics,
           University of T\"ubingen,
           D-72076 T\"ubingen, Germany}
}

\author{Michael Engelhardt}{
  address={Physics Department,
  New Mexico State University,
  Las Cruces, NM 88003, U.S.A.}
}

\author{Hugo Reinhardt}{
  address={Institute for Theoretical Physics,
           University of T\"ubingen,
           D-72076 T\"ubingen, Germany}
}

\begin{abstract}
In this talk, we review some recent results of the center vortex 
model for the infrared sector of $SU(3)$ Yang-Mills theory. Particular 
emphasis is put on the order of the finite-temperature deconfining phase 
transition and the geometrical structure of vortex branchings. 
We also present preliminary data for the 't Hooft loop operator and the 
dual string tension near the phase transition.
\end{abstract}

\maketitle

%%%%%%%%%%%%%%%%%%%%%%%%%%%%%%%%%%%%%%%%%%%%
%% MAINMATTER
%%%%%%%%%%%%%%%%%%%%%%%%%%%%%%%%%%%%%%%%%%%%

\subsection{Introduction}
The vortex picture of the Yang-Mills vacuum, initially proposed as a 
possible mechanism of colour confinement, has recently attracted 
a renewed attention. This is mainly due to the advent of new gauge 
fixing techniques which permit the detection of center vortex 
structures directly within lattice Yang-Mills configurations.
Numerical studies have revealed that the center 
projection vortices detected in this way do locate true physical objects 
(rather than lattice artifacts) \cite{R2}, and there is by now ample evidence 
that the infrared properties of Yang-Mills theory can be accounted for 
in terms of vortices \cite{R3}.

Based on these ideas, a random vortex world-surface model was introduced
as an effective low-energy description of $SU(2)$ 
Yang-Mills theory \cite{R4}; it has recently been extended to the gauge
group $SU(3)$ \cite{R5}. The fundamental assumption is that the long-range
structure of Yang-Mills theory is dominated by extended tubes of center flux 
tracing out closed surfaces in space-time. Consequently, we realise our model 
on a space-time lattice in which the fixed spacing $a$ represents the
transverse thickness of vortices. The random surfaces created on this 
lattice are weighted by a model action containing a Nambu-Goto and a curvature 
term, with dimensionless coupling constants $\epsilon$ and $c$, respectively. 
Physically, this means that vortices have a certain surface tension and they 
tend to be \emph{stiff}.
% The assignment of $Z(3)$ trialities to the elementary squares $q_{\mu\nu}(x)$ 
% in our model is such that large Wilson loops receive a center element 
% contribution for each linking with a (fat) vortex. 
For further details on our model and the determination of the parameters 
$\epsilon$ and $c$ (as well as the vortex extension $a=0.39\,\mathrm{fm}$), 
the reader is referred to \cite{R5}.

%%%%%%%%%%%%%%%%%%%%%%%%%%%%%%%%%%%%%%%%%%%%%%%%%%%%%%%%%%%%%%%%%%%%%%%%%%%%%

\subsection{Finite Temperature Phase Transition and Vortex Branching}

\begin{figure}
\includegraphics[height=.20\textheight]{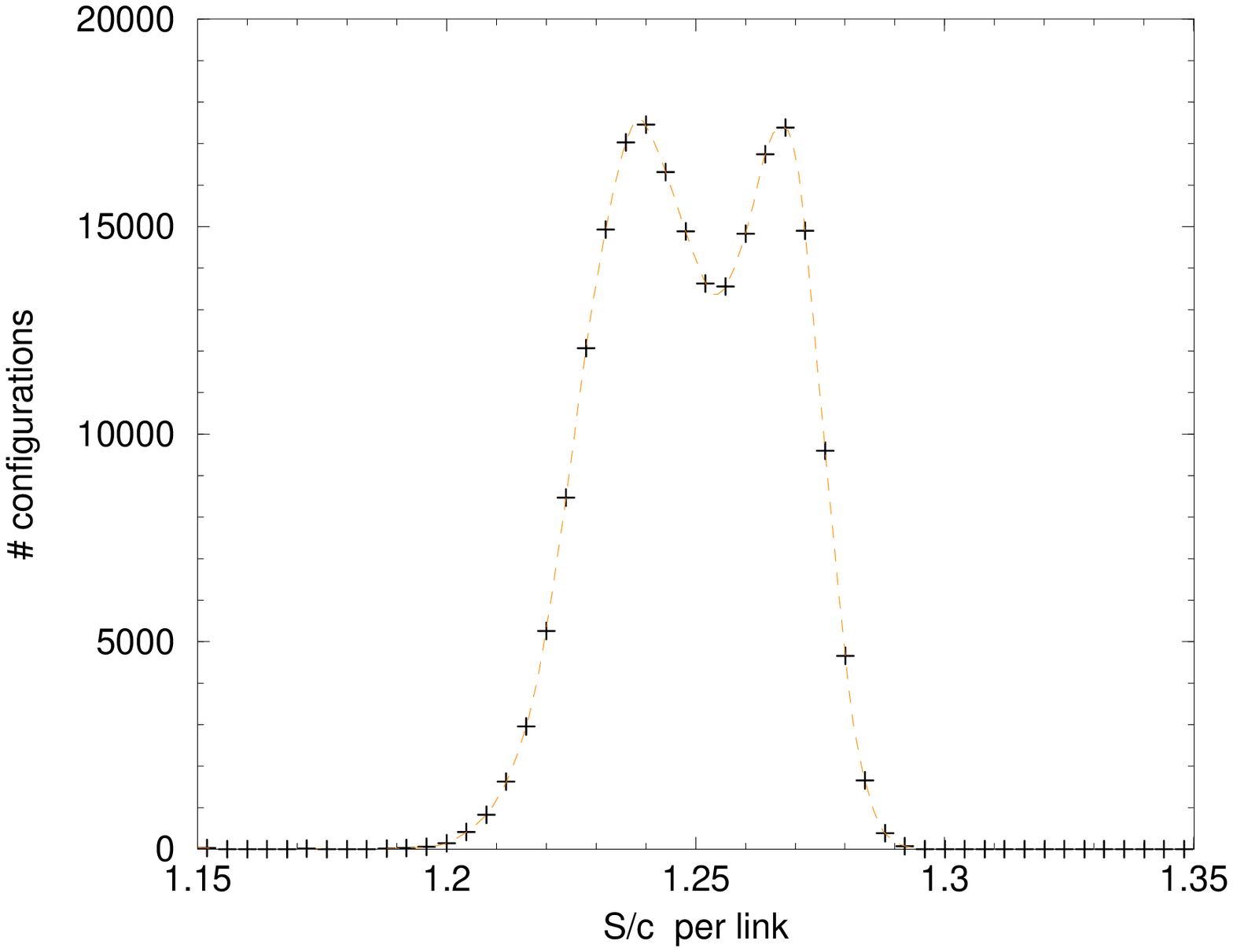}
\hspace{2cm}
\includegraphics[height=.20\textheight]{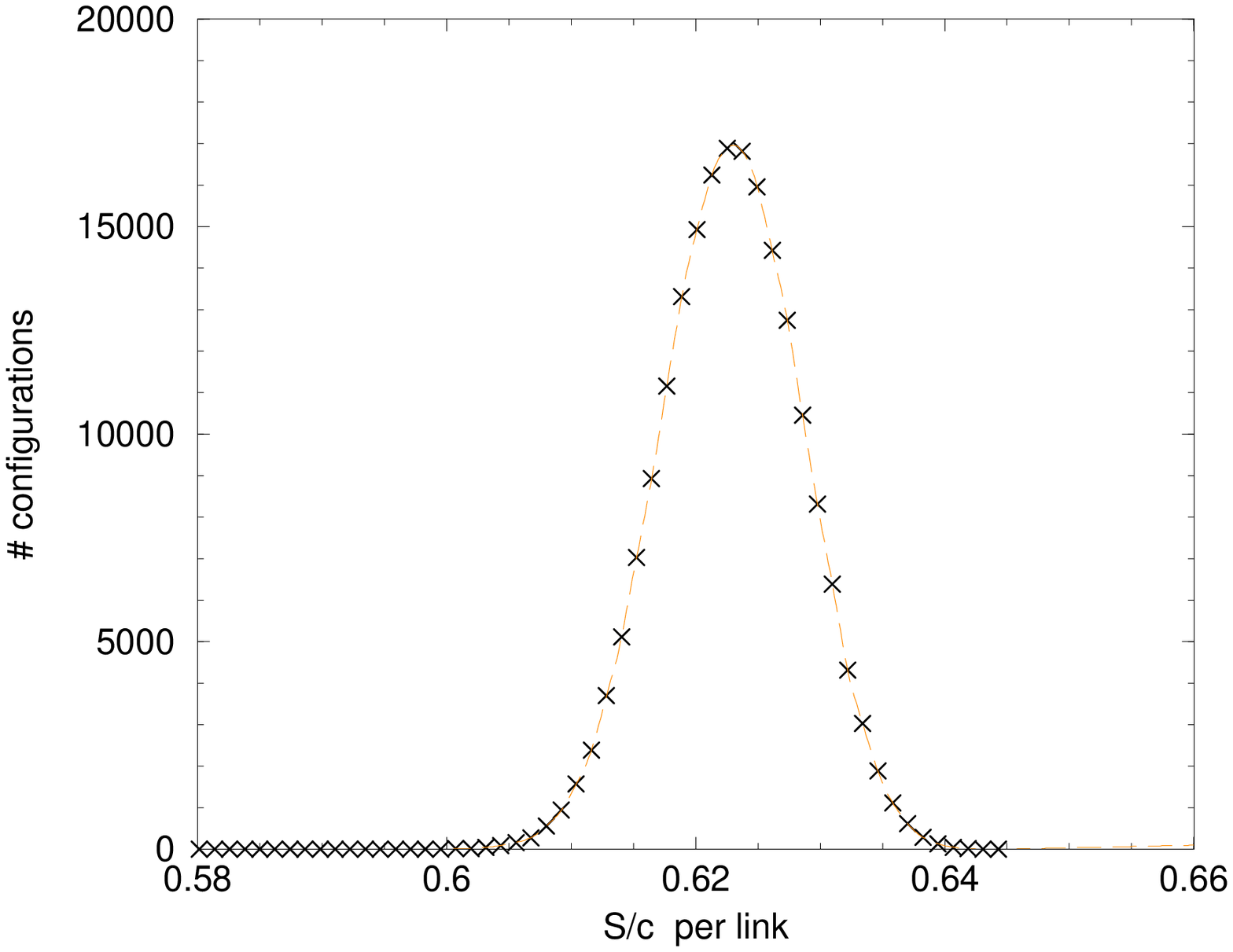}
\label{fig2}
\caption{Histograms of the average action density at the critical
point for $30^3 \times 2$ lattices; the left and right panel show 
$G=SU(3)$ and $G=SU(2)$, respectively.}
\end{figure}

Fig.~\ref{fig2} shows histograms of the action densities measured on 
$30^3 \times 2$ lattices at the critical points for the two gauge groups 
$G=SU(3)$ (left panel) and $G=SU(2)$ (right panel). As can be clearly seen, 
the $SU(3)$ transition exhibits the shallow double-peak structure 
characteristic for a weak \emph{first order transition}, while the $SU(2)$ 
transition is continuous (\emph{second order}). This 
qualitative behaviour is in agreement with findings from lattice 
gauge theory.

\begin{figure}
\includegraphics[height=.15\textheight]{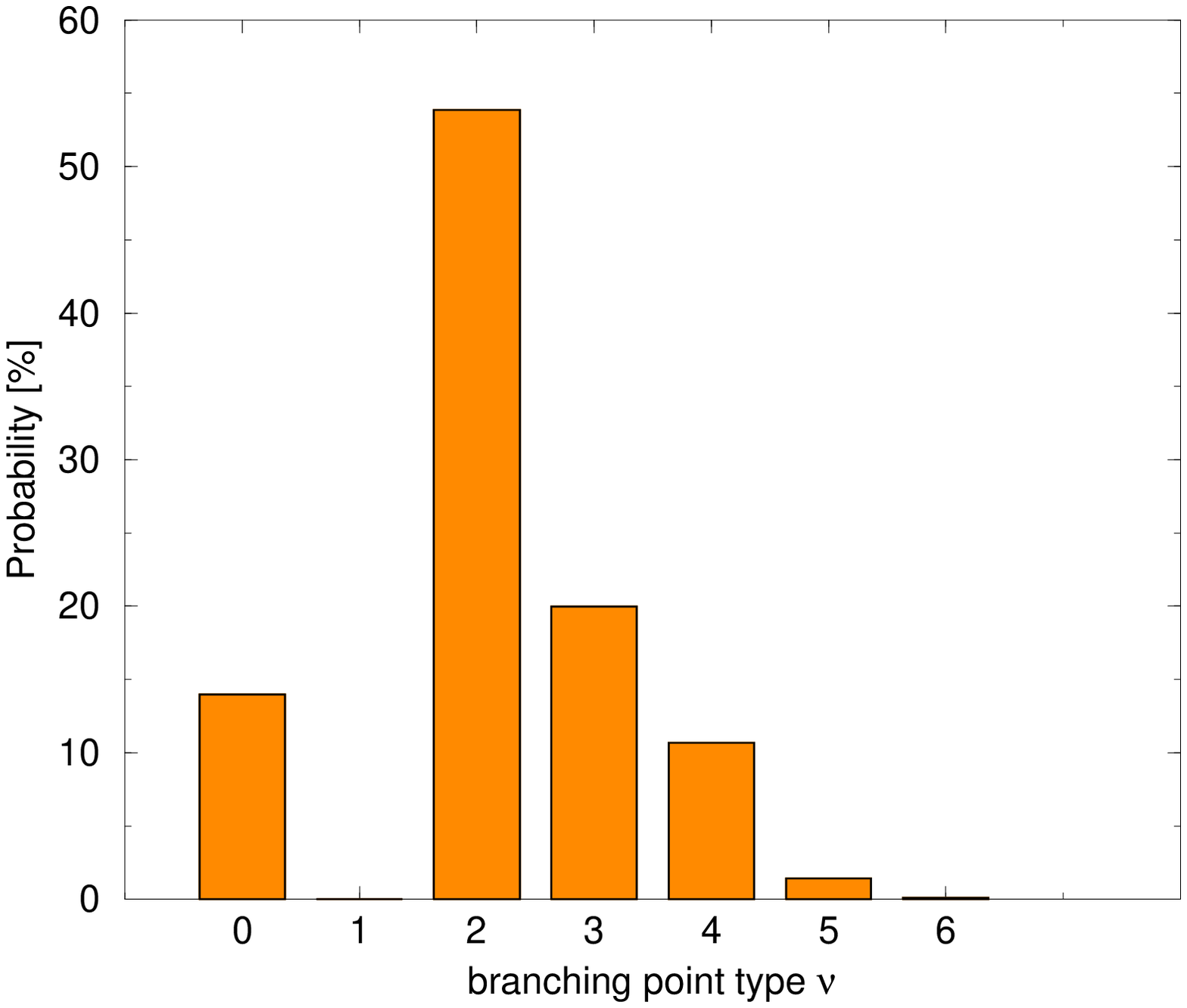}
\hspace{.5cm}
\includegraphics[height=.15\textheight]{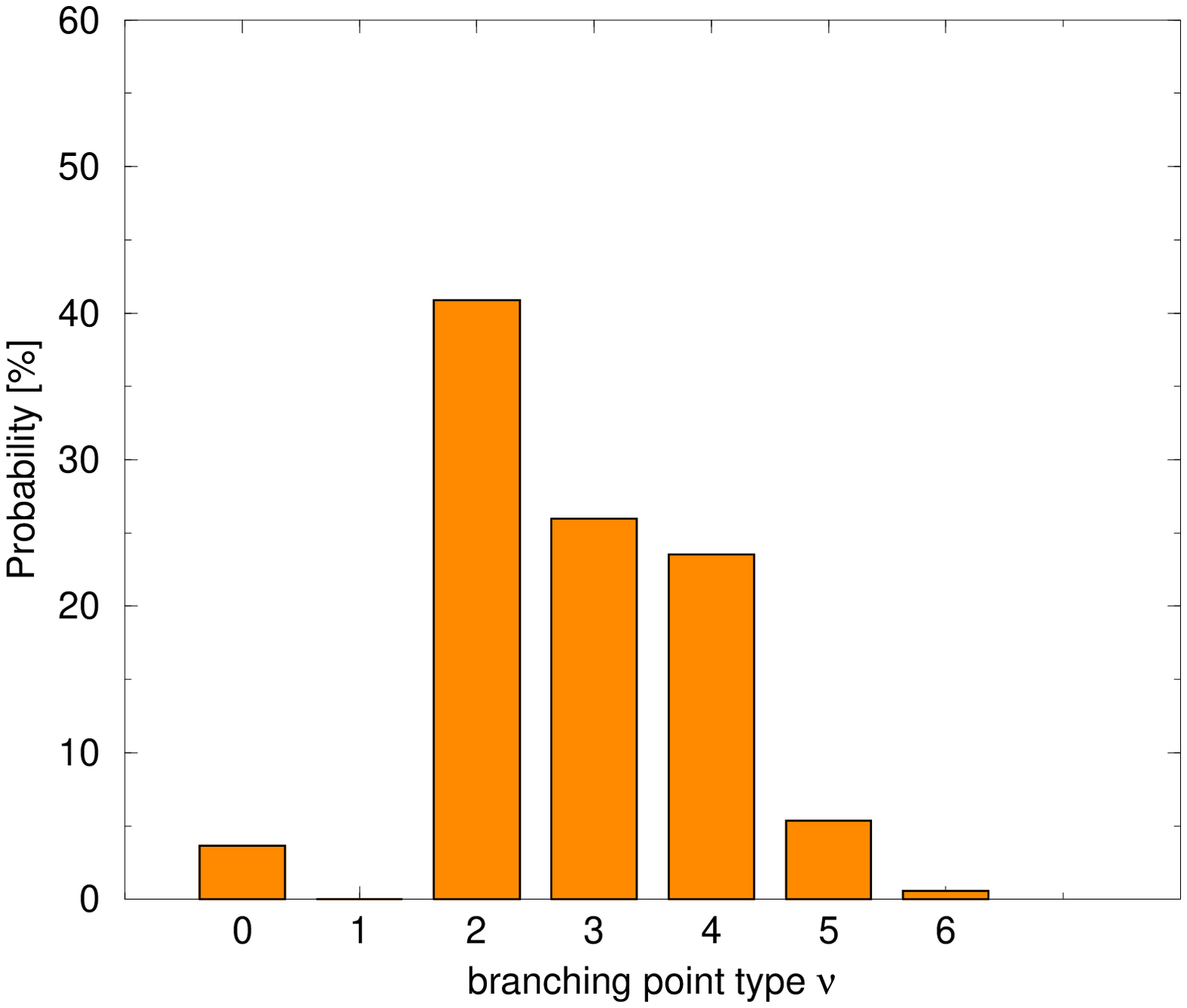}
\hspace{.5cm}
\includegraphics[height=.15\textheight]{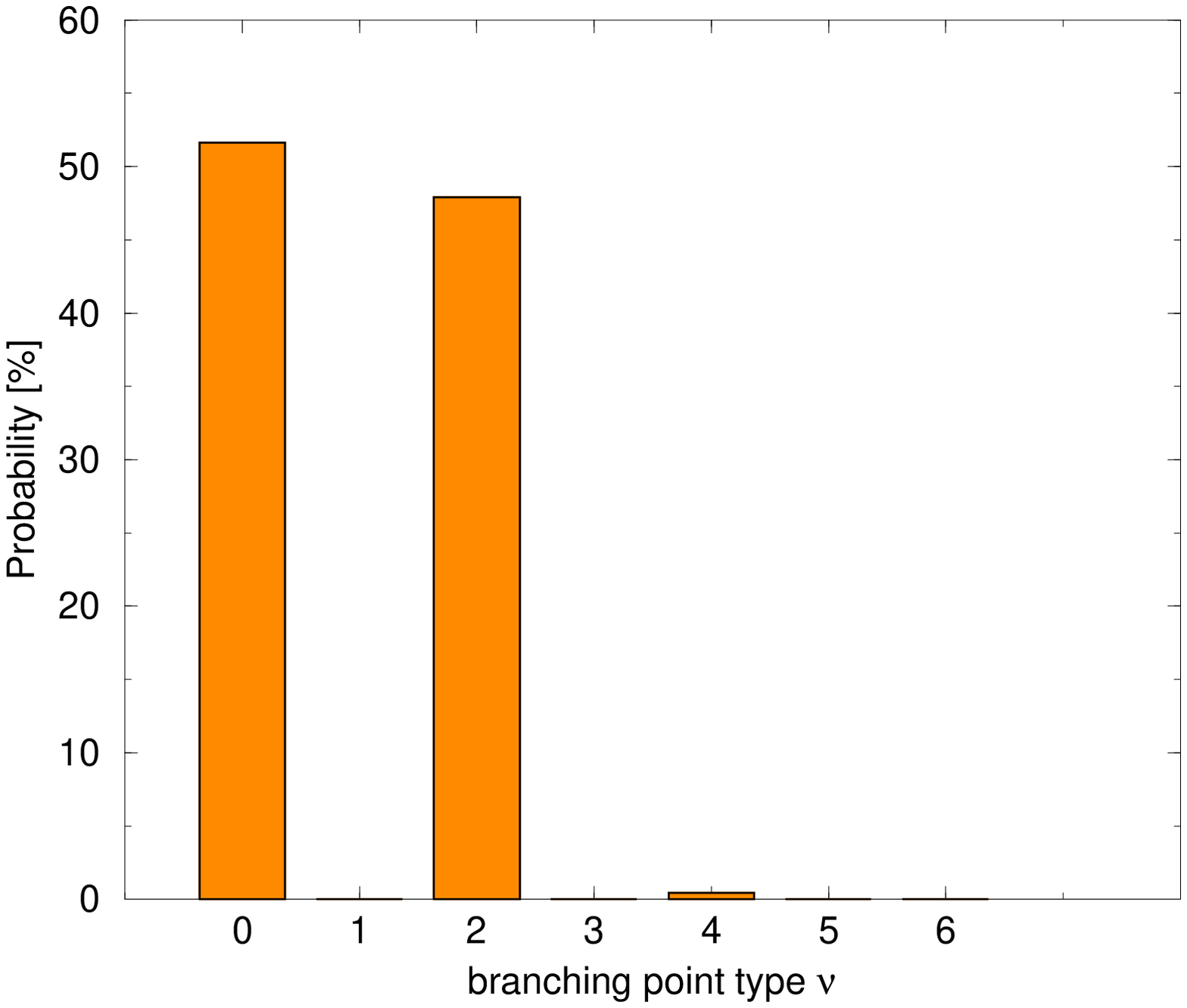}
\label{fig3}
\caption{Volume fractions occupied by points of a certain branching type 
$\nu$ within $3D$ lattice slices. The left panel shows the distribution at 
zero temperature ($c=0.21$, confined region). The middle and right panel 
both correspond to $T>T_c$, with the middle referring to a \emph{time slice} 
and the right to a \emph{space slice}.} 
\end{figure}

Since triality is only conserved $\emph{mod 3}$, an arbitrary number 
$\nu = 0,\ldots,6$ of vortex surfaces can meet at each link. The odd 
values $\nu = 3,5$ are not allowed in $SU(2)$ and represent 
genuine $SU(3)$ \emph{vortex branchings}. This phenomenon is best  
studied in $3D$ slices of the lattice, whence possible branching links are 
projected onto \emph{points} of type $\nu$. From fig.~\ref{fig3}, we 
conclude that the largest volume fraction in the confined phase  
corresponds to non-branching vortex matter ($\nu = 2$), with a considerable 
probability of both branchings ($\nu=3,5$) and self-intersections ($\nu=4,6$).
Only $15\,\%
$ of the volume is not occupied by vortices ($\nu=0$). 
In the deconfined phase ($T>T_c$), the situation is qualitatively unchanged 
for \emph{time-slices}, while \emph{space slices} show virtually 
zero branchings above $T_c$. This can be understood if the vortices undergo 
a \emph{(de)percolation phase transition} above $T_c$ and most
vortex clusters wind directly around the short time 
direction \cite{R4},\cite{R5}.  
\enlargethispage{1mm}
  
%%%%%%%%%%%%%%%%%%%%%%%%%%%%%%%%%%%%%%%%%%%%%%%%%%%%%%%%%%%%%%%%%%%%%%%%%%%%%

\subsection{'t Hooft Loop}

The 't Hooft loop can be viewed as a vortex creation operator \cite{R1}
that implements twisted boundary conditions when extended over an 
entire lattice plane \cite{R6}. It has been shown to be an alternative 
(dis)-order parameter for the deconfinement phase transition 
whose behaviour is \emph{dual} to the Wilson loop \cite{R6}.

\begin{figure}
\includegraphics[height=.20\textheight]{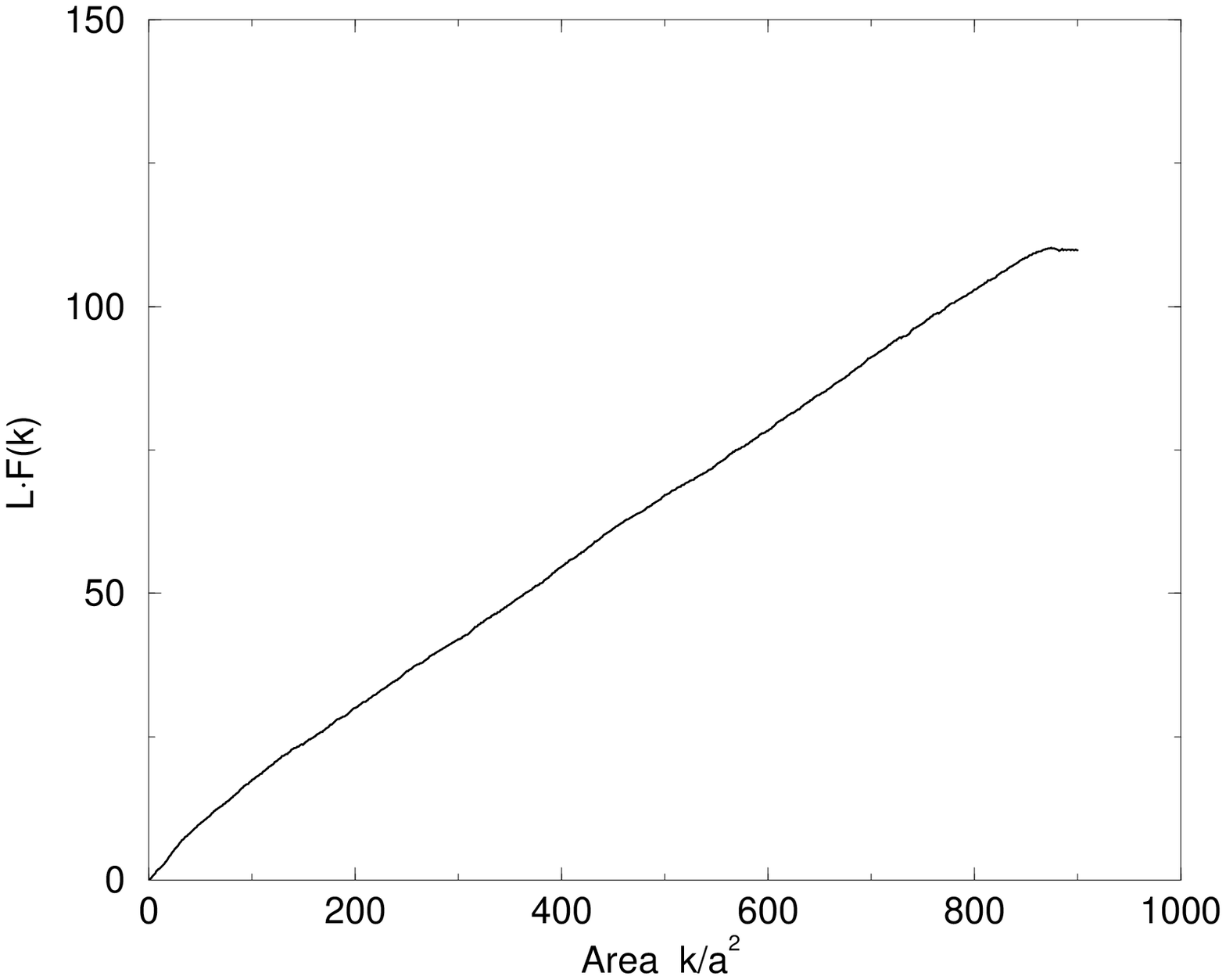}
\hspace{1cm}
\includegraphics[height=.20\textheight]{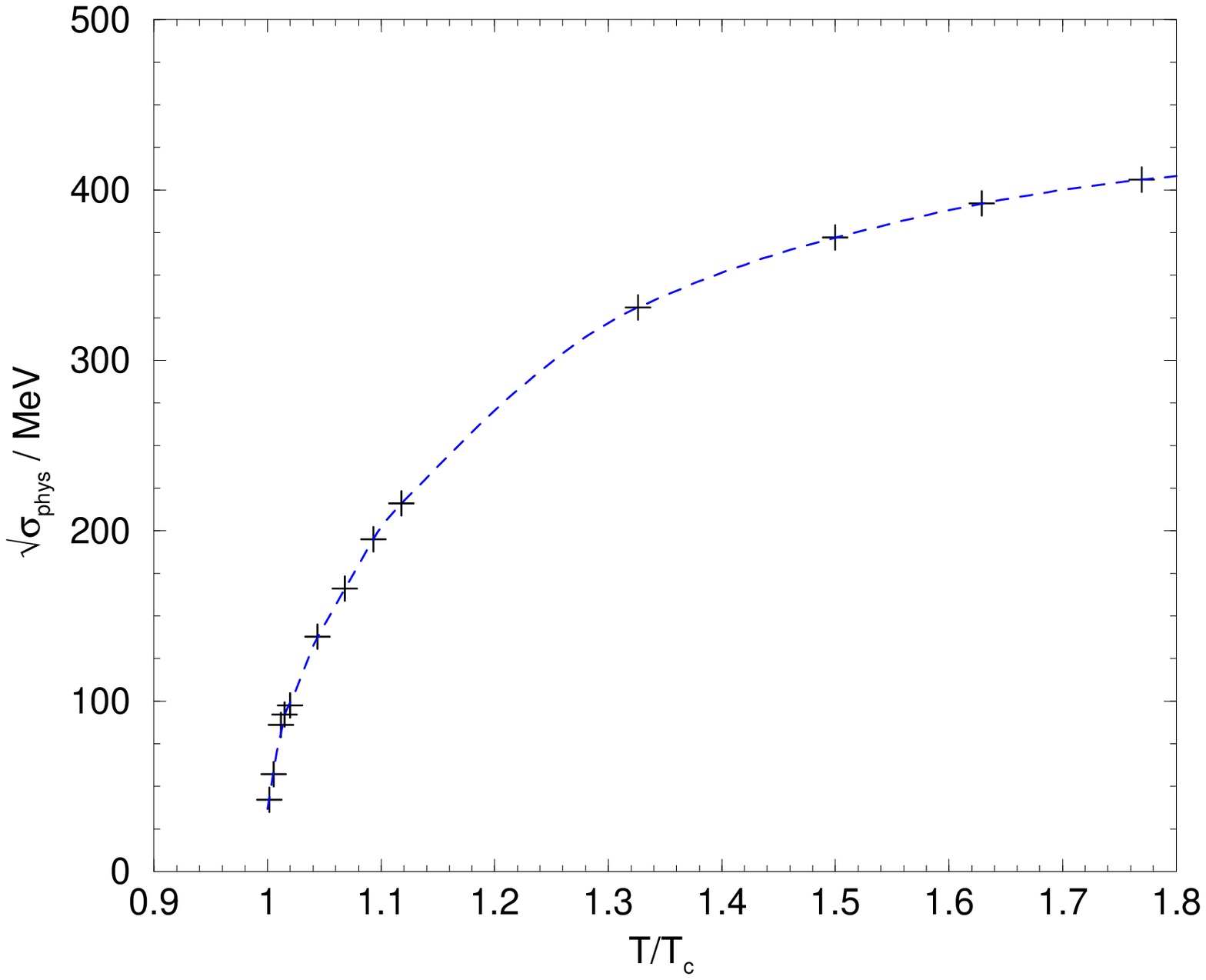}
\label{fig4}
\caption{Left Panel: Free energy of (incomplete) 't Hooft loops as 
a function of the loop area within the deconfined phase ($T/T_c = 1.093$). 
Right panel: The dual string tension $\tilde{\sigma}$ as a function of 
the temperature. Measurements were performed on a large 
$30^3 \times N_0$ lattice with $N_0 = 1,2$.}
\end{figure}

This expectation is confirmed in our model: The left panel of fig.~\ref{fig4} 
exhibits a linear rise of the free energy with the area of the 't Hooft loop,
% \footnote{The systematic deviations can be understood in terms of 
% $Z(3)$ monopole correlations.}
which permits to define a  \emph{dual string tension} in 
the deconfined phase. As we approach the phase transition from above, the 
dual string tension quickly vanishes (cf.~right panel of fig.~\ref{fig4}). 
Precise measurements close to the transition reveal a small discontinuity 
$\Delta \tilde{\sigma} \simeq (34\,\mathrm{MeV})^2$, which should be 
compared to the ordinary zero-temperature string tension 
$\sigma_0 = (440\,\mathrm{MeV})^2$ setting the overall scale. This 
demonstrates the weakness of the first order transition for $G=SU(3)$. 
  
\subsection{Conclusions}

In this talk, the physical foundation of the center vortex model for the 
infrared sector of $SU(3)$ Yang-Mills theory has been outlined. Only 
a selection of the results obtained so far could be presented. Among the 
effects discussed were the order of the deconfinement phase transition, 
the structure of branching points and the exact determination 
of the discontinuity in the free energy of the 't Hooft loop.
Interesting questions for future investigations are the study of 
deconfinement in higher colour groups, in particular the influence of 
complex branching patterns, as well as the coupling to 
quarks and the generation of a chiral condensate for $G=SU(3)$.

%%%%%%%%%%%%%%%%%%%%%%%%%%%%%%%%%%%%%%%%%%%%%%%%
%% BACKMATTER
%%%%%%%%%%%%%%%%%%%%%%%%%%%%%%%%%%%%%%%%%%%%%%%%

% \begin{theacknowledgments}
%   Grant $\ldots$ 
% \end{theacknowledgments}


\bibliographystyle{aipproc}   % if natbib is available

\end{document}